\documentclass[conference, a4paper]{IEEEtran}
\IEEEoverridecommandlockouts

\usepackage[font=footnotesize,caption=false]{subfig} 

\usepackage{cite}
\usepackage{amsmath,amssymb,amsfonts}
\usepackage{algorithmic}
\usepackage{graphicx}
\usepackage{textcomp}
\usepackage{xcolor}
\usepackage{booktabs}
\usepackage{glossaries}
\usepackage{subfig}
\usepackage{algorithm}
\usepackage{algorithmic}
\usepackage{bbm}
\usepackage{bm}
\usepackage{multirow}

\DeclareMathOperator*{\round}{round}
\DeclareMathOperator*{\argmax}{argmax}

\def\BibTeX{{\rm B\kern-.05em{\sc i\kern-.025em b}\kern-.08em
    T\kern-.1667em\lower.7ex\hbox{E}\kern-.125emX}}
\newacronym{3gpp}{3GPP}{3rd Generation Partnership Project}    
\newacronym{iot}{IoT}{Internet of Things}
\newacronym{ntn}{NTN}{Non-Terrestrial Network}
\newacronym{leo}{LEO}{Low Earth Orbit}
\newacronym{geo}{GEO}{Geosynchronous Earth Orbit}
\newacronym{isl}{ISL}{Inter-Satellite Link}
\newacronym{gsl}{GSL}{Ground-to-Satellite Link}
\newacronym{qos}{QoS}{Quality of Service}
\newacronym{ofdma}{OFDMA}{Orthogonal Frequency-Division Multiple Access}
\newacronym{b5g}{B5G}{5G and Beyond}
\newacronym{mimo}{MIMO}{Multiple-Input Multiple-Output}
\newacronym{embb}{eMBB}{Enhanced Mobile Broadband}
\newacronym{ue}{UE}{User Equipment}
\newacronym{nr}{NR}{New Radio}
\newacronym{gap}{GAP}{Generalized Assignment Problem}
\newacronym{mgap}{MGAP}{Multi-Level Generalized Assignment Problem}
\newacronym{csi}{CSI}{channel state information}    
\newacronym{ran}{RAN}{radio access network}
\newacronym{5g}{5G}{the 5th generation of mobile networks}
\newacronym{uav}{UAV}{unmanned aerial vehicle}
    
\begin{document}

\title{Continent-wide Efficient and Fair Downlink Resource Allocation in LEO Satellite Constellations}

\author{\IEEEauthorblockN{Israel Leyva-Mayorga\IEEEauthorrefmark{1}, Vineet Gala\IEEEauthorrefmark{2}, Federico Chiariotti\IEEEauthorrefmark{1}\IEEEauthorrefmark{3}, and Petar Popovski\IEEEauthorrefmark{1}}

\IEEEauthorblockA{\IEEEauthorrefmark{1}Department of Electronic Systems, Aalborg University, Denmark (\{ilm,fchi,petarp\}@es.aau.dk)}
\IEEEauthorblockA{\IEEEauthorrefmark{2} Department of Electrical Engineering, Indian Institute of Technology Bombay, India (vineetgala@ee.iitb.ac.in)}
\IEEEauthorblockA{\IEEEauthorrefmark{3} Department of Information Engineering, University of Padova, Italy (chiariot@dei.unipd.it)}}

\maketitle

\begin{abstract}
 The integration of \gls{leo} satellite constellations into 5G and Beyond is essential to achieve efficient global connectivity. As \gls{leo} satellites are a global infrastructure with predictable dynamics, a pre-planned fair and load-balanced allocation of the radio resources to provide efficient downlink connectivity over large areas is an achievable goal. In this paper, we propose a distributed and a global optimal algorithm for satellite-to-cell resource allocation with multiple beams. These algorithms aim to achieve a fair allocation of time-frequency resources and beams to the cells based on the number of users in connected mode (i.e., registered). Our analyses focus on evaluating the trade-offs between average per-user throughput, fairness, number of cell handovers, and computational complexity in a downlink scenario with fixed cells, where the number of users is extracted from a population map. Our results show that both algorithms achieve a similar average per-user throughput. However, the global optimal algorithm achieves a fairness index over $0.9$ in all cases, which is more than twice that of the distributed algorithm. Furthermore, by correctly setting the handover cost parameter, the number of handovers can be effectively reduced by more than $70\%$ with respect to the case where the handover cost is not considered.
\end{abstract}
\section{Introduction}
\label{sec:intro}
Reducing inequality among regions is one of the 17 Social Development Goals (SDGs) defined by the United Nations (UN). An essential milestone towards this ambitious goal is, as evidenced by the European Commission digital strategy, to achieve ubiquitous broadband connectivity across entire continents~\cite{EC}. Deploying a resilient and widespread infrastructure for \gls{5g} in densely populated areas, such as large and highly-developed cities is attainable both economically and geographically, but these areas are already well-served by a combination of \gls{5g}, 4G, WiFi, and cabled infrastructure. In contrast, one of the major challenges for \gls{5g} is to provide resilient broadband and \gls{iot} connectivity in remote and rural areas~\cite{Yaacoub2020, Simonsen22, xie2021}. 

However, terrestrial infrastructure alone cannot bring \gls{5g} connectivity to rural and remote communities. Therefore, the development of architectures and mechanisms for the integration of \gls{ntn} into 5G has been a hot research topic for several years now~\cite{Guidotti2018}. Satellites deployed at \gls{leo}, between $500$ and $2000$\,km above the Earth's surface, are of particular interest, as their relatively low altitude of deployment results in a typical propagation delay from ground to satellite as low as $2$\,ms, which is sufficient to serve a wide range of low-latency applications~\cite{Leyva-Mayorga2020}. 

With Release 17, the long-awaited integration of \gls{ntn} into \gls{5g} \gls{nr} is now a reality~\cite{TS38.300}, covering the essential \gls{ran} mechanisms and procedures to communicate ground users directly with \gls{leo} satellites, which include signaling, mobility management, and handovers~\cite{3GPPTR38.821}. The architecture considered in this first integration is bent-pipe, where the \gls{leo} satellites serve as relays towards an \gls{ntn} gateway. Moreover, two options for user-to-cell mapping with \gls{leo} are considered, namely, Earth-moving cells and quasi-Earth-fixed cells~\cite{TS38.300}. Earth-moving cells follow the nadir point of \gls{leo} satellites as they orbit the Earth and, hence, require continuous mobility management, in the form of handovers, for the users that enter and leave the cells. In contrast, quasi-Earth-fixed cells cover a fixed geographical area, which must be followed by the satellites throughout the pass via beam steering. While beam steering can be achieved with traditional spot beams, advances in massive \gls{mimo} for \gls{leo} satellites might enable to achieve full frequency reuse in the future~\cite{You2020}

By maintaining the user-to-cell mapping, the use of quasi-Earth-fixed cells eliminates the need for continuous handovers. Instead, all the users in a cell must perform  handover at a specific point in time. While naively performing a large number of handovers simultaneously might create a signaling peak that can overload the control channels of \gls{5g} \gls{nr}, an ephemeris can be used to accurately predict the movement of the satellites and to avoid congestion by preparing the handovers in advance~\cite{TS38.300, Juan2022}.

Downlink resource allocation over wide areas with \gls{leo} satellite constellations is a complex problem that involves numerous satellites and a large number of users. Furthermore, allocating resources with the sole objective of maximizing efficiency might exacerbate the already-existing inequalities among geographical regions. Therefore, achieving an adequate balance between (proportional) fairness and overall efficiency in resource allocation is highly desirable. Hence, fairness and efficiency trade-offs have been widely studied in  multi-agent systems with reinforcement learning, namely, to define reward functions that promote fair and efficient solutions~\cite{Jiang2019}. Multi-armed bandit (MAB) approaches have been proposed to achieve a fair allocation of resources relatively small areas with a small number of satellites, \glspl{uav}, and users~\cite{Arani2021}. However, the approach in~\cite{Arani2021} relied on having an accurate mobility model to optimize the trajectories of the \glspl{uav}, which limits its applicability. Furthermore, the fairness index calculated in~\cite{Arani2021} was observed to decrease with the number of users, even when the evaluation scenario considered at most $300$ users.

In this paper, we study a \gls{5g} and beyond downlink scenario with direct user-to-\gls{leo} satellite access over wide areas with quasi-Earth fixed cells based on real population data~\cite{CIESIN}. We propose two algorithms for proportional fair resource allocation that consider the cost of handovers and that aim to provide uniform connectivity based on the number of users in connected mode within the area. The handover cost aims to reduce the number of handovers by reflecting the loss in the overall network efficiency due to the execution of the handover procedure, but also to higher layer mechanisms, such as flow and congestion control, that are affected by the link, latency, and capacity changes in the routes~\cite{xie2021}. To the best of our knowledge, this is the first study that focuses on fair and efficient resource allocation over wide areas that considers the handover cost and real-life geographical user distribution.

\begin{figure}[t]
\centering
\subfloat[Cell-level population map of the considered region.]{\includegraphics[width=0.85\columnwidth]{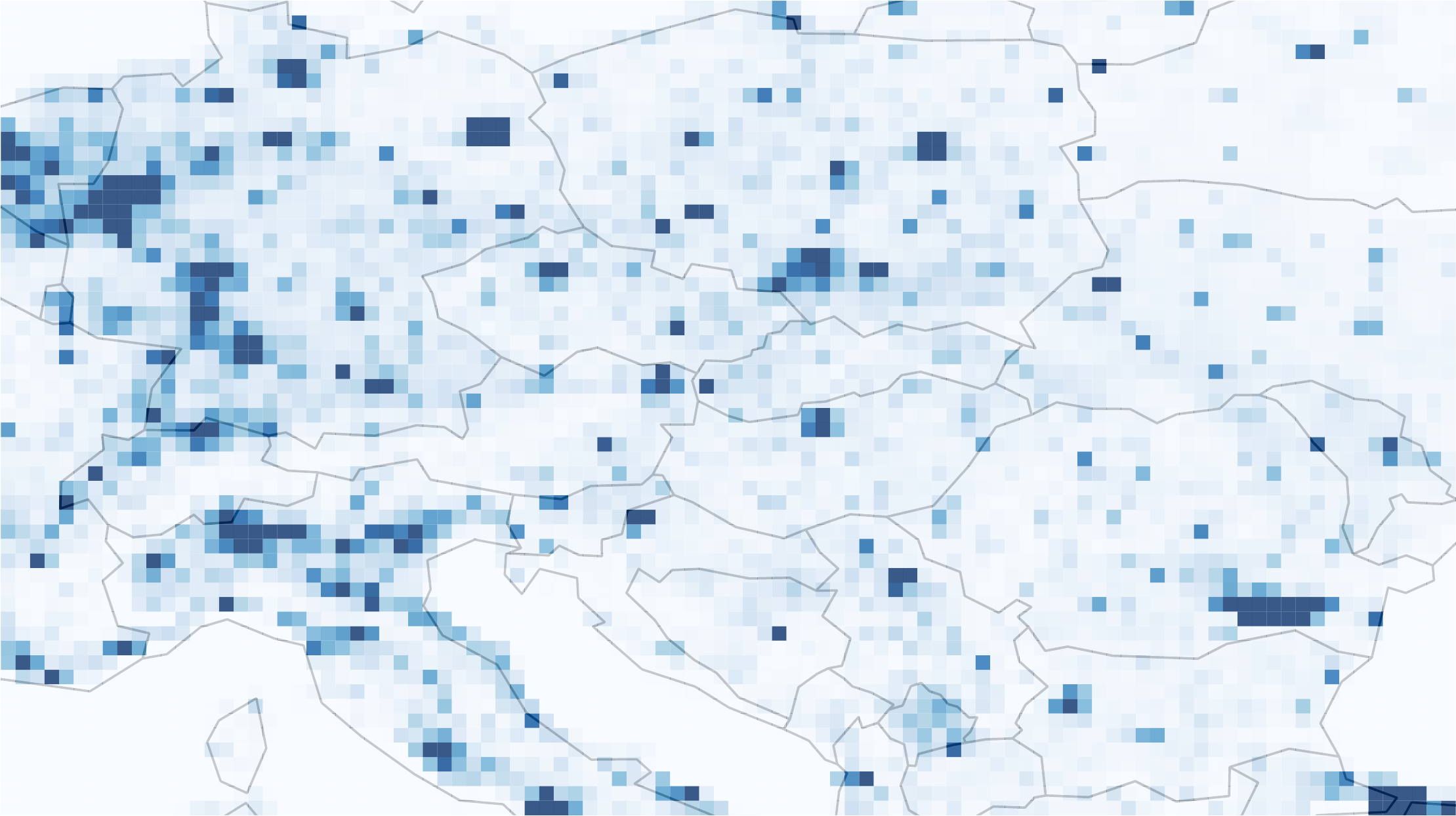}\label{fig:users_map}}\\
\subfloat[Throughput map with distributed fair resource allocation.]{\includegraphics[width=0.85\columnwidth]{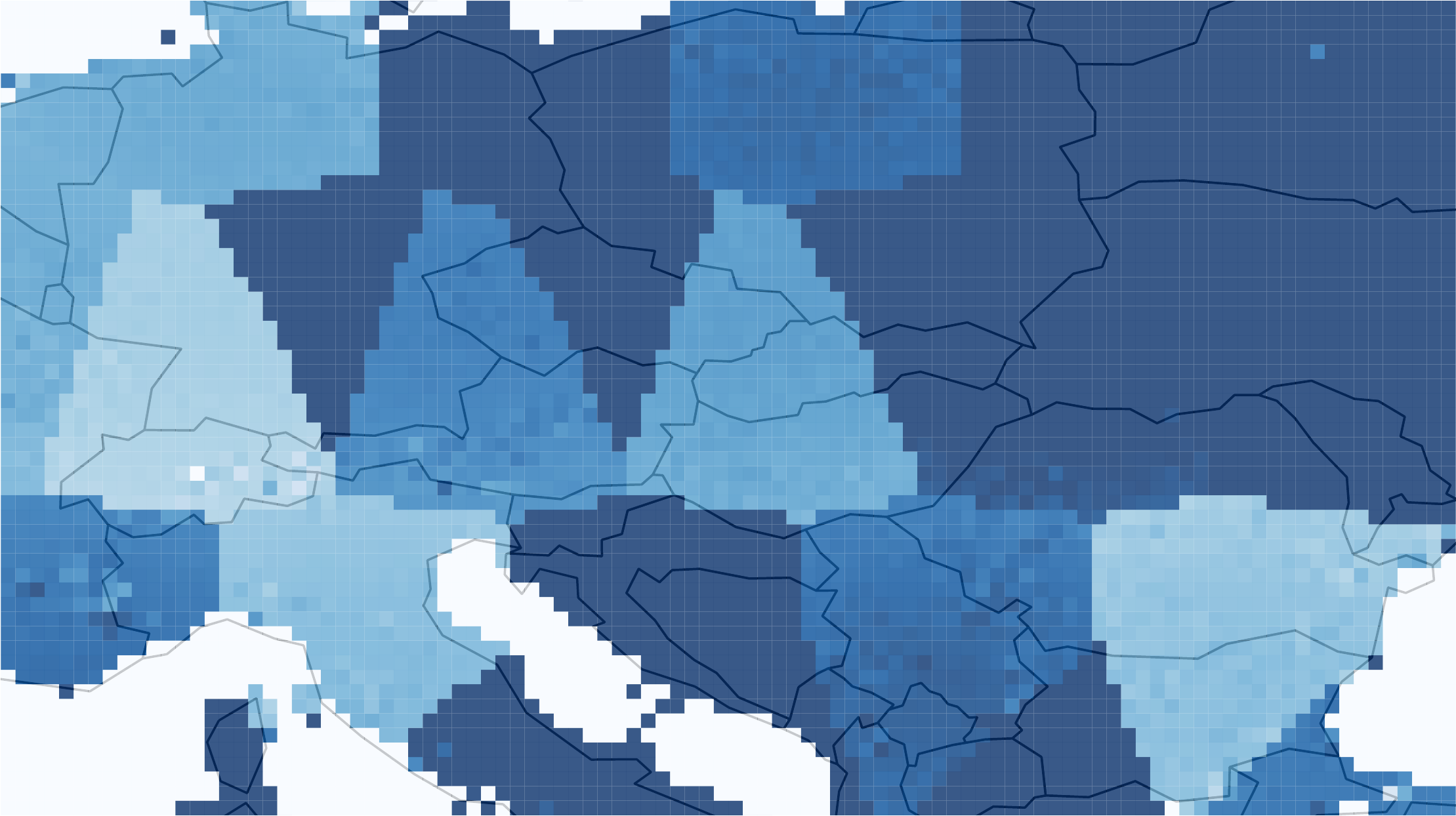}\label{fig:distributed_rates_map}}\\
\subfloat[Throughput map with global optimal fair resource allocation.]{\includegraphics[width=0.85\columnwidth]{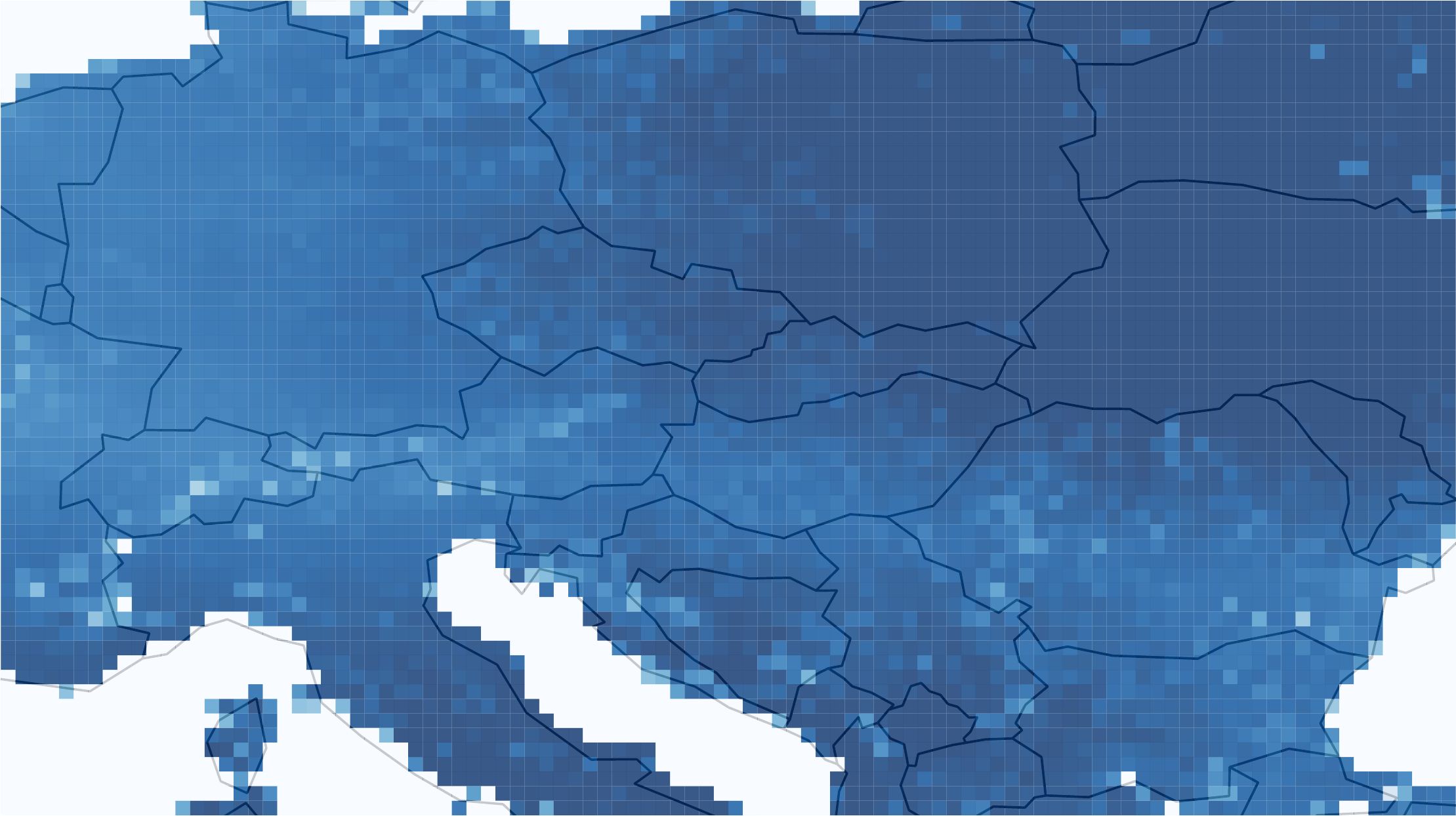}\label{fig:optimal_rates_map}}
\caption{Map of central Europe illustrating the proportional-fair resource allocation problem. The number of active users per cell is $0.1\%$ of the total population~\cite{CIESIN}. The same throughput ranges were set for (b) and (c).}
\label{fig:map}
\end{figure}

The effects of our proposed solutions can be seen at a glance from Fig.~\ref{fig:map}: as the population is unevenly distributed among cells, a naive allocation based on matching the closest satellite to each cell leads to significant unfairness, even when redistributing each satellite's resources. On the other hand, a joint resource allocation and cell-to-satellite mapping can achieve much higher fairness, and, as we will discuss in the paper, the same or better overall throughput with fewer handovers as well.

The rest of the paper is organized as follows: first, the general system model is presented in Sec.~\ref{sec:sys_model}. Sec.~\ref{sec:problem} then describes the resource allocation problem and the two proposed algorithms, which are evaluated by simulation in Sec.~\ref{sec:results}. Finally, Sec.~\ref{sec:conc} concludes the paper and presents some possible avenues of future work.

\section{System Model}\label{sec:sys_model}
Let us consider a downlink \gls{ntn} scenario in which a \gls{leo} satellite constellation with $S$ satellites serves users on the ground through a direct user access link~\cite{TS38.300,Guidotti2018}. We limit our analysis to a fixed geographical region of the Earth. Following the fixed cell scenario described by the \gls{3gpp}~\cite{3GPPTR38.821, TS38.300}, the users in the region are aggregated into $C$ fixed and uniformly distributed geographical cells as shown in Fig.~\ref{fig:users_map}. The set of cells is denoted as $\mathcal{C}$, the area of a cell $c\in\mathcal{C}$ is denoted as $A_c$, and the total number of users in the cell is $U_c^\text{tot}$. A fraction $\alpha_c\in\left[0,1\right]$ of the users in cell $c$ are actively receiving data from the satellite network and $1-\alpha_c$ of the users are either inactive or communicate through the terrestrial infrastructure. Hence, only $U_c=\alpha_c U_c^\text{tot}$ users in cell $c\in\mathcal{C}$ are actively receiving data in the downlink from the satellite network.

We consider a multi-beam \gls{ofdma} system similar to one described by the \gls{3gpp} for \gls{ntn} 5G \gls{nr}, with $N_B$ beams per satellite and a total bandwidth per beam of $B$\,Hz. Each \gls{ofdma} frame occupies $T_F$\,seconds in the time domain. 
The satellites are equipped with high gain antennas that allow them to provide high data rates to the users at ground despite the high path loss from \gls{leo}. Furthermore, the satellites possess precise beam steering capabilities, which allos them to point the beams toward the intended cells throughout the satellite pass. In addition, we consider that the satellites can steer each of the $N_B$ beams once per \gls{ofdma} frame, either to align them to the same cell or to cover a different cell.

We analyze a time frame divided into $K$ time slots of duration $T$, such that $kT,\ k\in\left\{0,1,\dotsc,K-1\right\}$ is the start time of the $k$-th time slot. The set of satellites in the constellation is denoted as $\mathcal{S}$ and the set of those within range of any cell $c\in\mathcal{C}$ at time slot $k$ is $\mathcal{S}_k$. The allocation of resources from the satellites $s\in\mathcal{S}_k$ to the cells can only be changed at the beginning of each time slot. 
In the time domain, each time slot contains $N_T = T/T_F$ \gls{ofdma} frames and the minimum unit for resource allocation is one frame. In the frequency domain, the whole bandwidth per beam $B$ is used.

We consider the worst-case performance for a user in a cell $c$. Therefore, the attenuation of the signal between a cell $c$ and a satellite $s$ at time $kt$ is given by the free space path loss, the atmospheric attenuation $\theta$, and the pointing loss $\ell$ as
\begin{equation}
   \mathcal{L}_k(s,c) = \left(4 \pi d_k(s,c) f\right)^2 v_c^{-2} \theta\ell.
\end{equation}
where $d_k(s,c)$ is the maximum distance between satellite $s$ and any point in cell $c$ at time $kt$,  $v_c$ is the speed of light, and $f$ is the carrier frequency used for communication.

Let $P^{(s)}$ be the transmission power from the satellite, $G^{(c,s)}$ and $G^{(s,c)}$ be the antenna gain of the users in cell $c$ towards satellite $s$, and vice versa, respectively, and $\sigma^2$ is the noise power at the users. The nominal data rate for a satellite-cell pair $(s,c)$ at the beginning of slot $k$ is upper-bounded by:
\begin{equation}
    \rho_k(s,c) = B\log_2 \left( 1+ \frac{P^{(s)}G^{(c,s)}G^{(s,c)}}{\mathcal{L}_k(s,c)\sigma^2} \right).
\end{equation}
However, considering only the instantaneous achievable rate may lead to outages within the slot due to the satellite moving out of the coverage area of a cell $c$ or to improper rate selection. Therefore, the minimum  nominal rate at the edges of the matching interval is considered, which is defined as 
\begin{equation}
    \rho_k^\text{min}(s,c) = \min\left(\rho_k(s,c), \rho_{k+1}(s,c)\right).
\end{equation}

We assume that the \gls{ofdma} frames, along with the multiple resource blocks within each \gls{ofdma} frame, belonging to a same cell and satellite beam pair are allocated uniformly to the users within the cell.

\section{Resource Optimization}
\label{sec:problem}
We now formulate an optimization problem to allocate satellite beams and their \gls{ofdma} resources to cells. 
Let $\mathbf{X}_k(s,c)\in\{0,\ldots,N_TN_B\}$ be the number of \gls{ofdma} frames allocated to cell $c$ by satellite $s$ in slot $k$, and $\mathbf{X}_k\in\{0,\ldots,N_TN_F\}^{S\times C}$ be the resource allocation matrix at time slot $k$, where its $(s,c)$-th element is $\mathbf{X}_k(s,c)$. Furthermore, let $\mathbf{x}_k^{(s)}$ be the allocation vector for satellite $s$ at time slot $k$, where its $c$-th element is $\mathbf{X}_k(s,c)$. Building on this, we define the minimum throughput for any user in cell $c \in \mathcal{C}$ served by satellite $s$ within slot $k$ for a given $\mathbf{X}_k(s,c)$ and $x_{k-1}(s,c)$ as
\begin{equation}
    R_k\left(s,c\right) = \frac{T_F}{TU_c} \mathbf{X}_k(s,c) \rho_k^\text{min}(s,c).
    \label{eq:rate}
\end{equation}

Finally, let  $h_{\text{cost}}\in\left[0,1\right)$ be the handover cost, representing the impact of switching a cell from one satellite to another on higher layer mechanisms~\cite{xie2021}. Next, let $H(x)$ be the Heaviside step function, equal to $1$ if $x>0$ and $0$ otherwise. Building on these, we define the handover penalty for time slot $k$ based on the allocation at the previous time slot $k-1$ as
\begin{equation}
    h_k(s,c)=h_{\text{cost}}\left(1-H(x_{k-1}(s,c))\right), 
\end{equation}
so that a higher value of $h_{\text{cost}}$ increases the handover penalty.

Our goal is to determine the values of $\mathbf{X}_k$, the allocation of satellites and their resources to the cells, so that the resources are efficiently and fairly distributed across the cells in the area to achieve a uniform coverage while minimizing handovers. This is modeled as a load balancing problem, where the nominal data rate of individual individual cells is maximized cells when being served by the closest satellite. However, due to the uneven geographical distribution of users within the cells, it might be convenient to connect to a satellite that is farther away, but that has more available resources or that eliminates the need for a handover.

The following three constraints are defined for the allocation within one time slot. First, one satellite can allocate up to $N_T$ \gls{ofdma} frames to each cell $c\in \mathcal{C}$. Second, each satellite cannot allocate more than $N_TN_B$ resources to the cells. Third, a cell $c \in \mathcal{C}$ can never be served by multiple satellites. Hence, up to one satellite can allocate between $1$ and $N_T$ \gls{ofdma} frames to a given cell $c$ and the rest of the satellites must allocate $0$ frames to the cell. With these constraints in place, we formulate the optimization problem as follows.
\begin{IEEEeqnarray*}{rll}
\mathbf{X}_k^*=\,&\argmax_{\mathbf{X}_k}&\sum_{c\in\mathcal{C}}U_c\log \Big(\sum_{s\in\mathcal{S}}R_k(s,c)\big(1-h_k(s,c)\big)\Big),\IEEEyesnumber\\
&\text{subject to}~&
\mathbf{X}_k(s,c)\in\{0,1,\dotsc, N_T\},  \forall (s,c) \in \mathcal{S}\times\mathcal{C},\\[0.5em]
&&\displaystyle \sum_{c \in \mathcal{C}}\mathbf{X}_k(s,c)\in\{0,1,\dotsc, N_TN_B\},\  \forall s \in \mathcal{S},\\
&&\displaystyle\sum_{s \in \mathcal{S}} H(\mathbf{X}_k(s,c))\in\{0,1\}, \ \forall c \in \mathcal{C},
\label{eq:problem_x}
\end{IEEEeqnarray*}

It is trivial to see that the problem requires solving the multiple knapsack problem, which is NP-hard. No efficient optimization algorithm exists for this class of problems, but the optimal solution can be closely approximated by \emph{(i)}, relaxing the first and second constraints to accommodate a real-valued optimization variable $\hat{\mathbf{X}}_k\in\left[0,N_T\right]^{S\times C}$, and \emph{(ii)}, reformulating the optimization objective to include a set of weighting terms $w(s,c)$ that promote the sparsity of the real-valued optimization variable $\hat{\mathbf{X}}_k$ and, hence, allow us to remove the third constraint as follows.
\begin{IEEEeqnarray*}{rll}
\hat{\mathbf{X}}_k^*=\,& \argmax_{\hat{\mathbf{X}}_k}& \hspace{-0.8em}\sum_{c\in\mathcal{C}}\!\Bigg(U_c\log \Big(\sum_{s\in\mathcal{S}}\hat{R}_k(s,c)\big(1-h_k(s,c)\big)\Big)\IEEEeqnarraynumspace\\[-0.2em]
&&\IEEEeqnarraymulticol{1}{r}{-\!\sum_{s\in\mathcal{S}}w(s,c)\hat{x}_k(s,c)\Bigg)},\IEEEeqnarraynumspace\IEEEyesnumber\label{eq:problem_cvx}\\[0.2em]
&\text{subject to}~&
\displaystyle0\leq \hat{x}_k(s,c)\leq N_T, \quad\forall s \in \mathcal{S}, c\in\mathcal{C},\\[0.2em]
&&\displaystyle \sum_{c \in \mathcal{C}}\hat{x}_k(s,c)\leq N_TN_B, \quad\forall s \in \mathcal{S},
\end{IEEEeqnarray*}
where $\hat{R}_k(s,c)$ is obtained by substituting $\mathbf{X}_k(s,c)$ with $\hat{x}_k(s,c)$ in~\eqref{eq:rate}.

\begin{algorithm}[t]
\caption{Resource allocation adjustment.}
\label{alg:valid_allocation}
\begin{algorithmic}[1]
\footnotesize
\REQUIRE $\mathbf{X}_{k}$
\FOR {$c\in\mathcal{C}$}
\IF {$\sum_{s\in\mathcal{S}}H(x_k(s,c))>1$}
\STATE $s^*\leftarrow \argmax_s \mathbf{X}_k(s,c)\rho^\text{min}_k(s,c)h_k(s,c)$
\STATE $\mathbf{X}_k(s,c)\leftarrow 0$ for all $s\neq s^*$
\ENDIF
\ENDFOR
\FOR {$s\in\mathcal{S}$}
\WHILE{$\sum_{c\in\mathcal{C}}\mathbf{X}_k(s,c)>N_TN_B$}
\STATE $c'\leftarrow \argmax_c \mathbf{X}_k(s,c)-\hat{x}_k(s,c)$
\STATE $\mathbf{X}_k(s,c')\leftarrow \mathbf{X}_k(s,c')-1$
\ENDWHILE
\ENDFOR
\RETURN $\mathbf{X}_k$
\end{algorithmic}
\end{algorithm}

The problem in \eqref{eq:problem_cvx} and its constraints are convex and can be solved efficiently using a weighted $\ell_1$ heuristic. The latter is an iterative method, where the weights $w(s,c)$ are first initialized to $0$. Then, at each iteration, the convex problem is solved, for example, using interior point methods or Lagrangian dual methods. Then, the weights are updated as $w(s,c)=\beta/\left(\tau+\hat{x}(s,c)\right)$, which serve as a penalty for allocating resources from satellite $s$ to cell $c$. After several iterations, the values of $\hat{\mathbf{X}}_k(s,c)$ will converge to an optimal solution $\hat{\mathbf{X}}_k^*$. Afterwards, the real-valued optimal allocation $\hat{\mathbf{X}}_k^*$ can be mapped to discrete values to obtain the final solution $\mathbf{X}^*_k$, which must fulfill the constraints set in~\eqref{eq:problem_cvx}. A common solution for this latter step is to simply apply the standard rounding function to $\hat{\mathbf{X}}_k(s,c)$. Afterwards,  Algorithm~\ref{alg:valid_allocation} is executed to ensure that the allocation is discrete and respects the constraints. 

Furthermore, since the weights $w(s,c)$ are initialized to $0$, it is likely that some cells are allocated resources from more than one satellite in the first few iterations. These are hereafter called \emph{conflicting cells}. That is, it cannot be guaranteed that there will be no conflicting cells if the number of iterations $n_\text{iter}$ is low. In such cases, the solution obtained with Algorithm~\ref{alg:el1} is infeasible according to the last constraint in~\eqref{eq:problem_x}. On the other hand, the number of iterations needed to ensure that the solution is feasible might be exceedingly large and, thus, the execution time of the global optimal algorithm might become prohibitively large.
To avoid these problems, the simple procedure  shown in Algorithm~\ref{alg:valid_allocation} is used after rounding to modify the allocation $\hat{\mathbf{X}}_k(s,c)$ so that it finds the closest point to $\hat{\mathbf{X}}_k(s,c)$ in the feasible region of the fully constrained optimization problem~\eqref{eq:problem_x}.

{\renewcommand{\arraystretch}{1}
\begin{algorithm}[t]
\caption{Weighted $\ell_1$ algorithm for global optimization.}\label{alg:el1}
\begin{algorithmic}[1]
\footnotesize
\REQUIRE $\mathbf{X}_{k-1}$, $\beta, \tau$, and $\rho_k^\text{min}(s,c)$ for all $s,c$
\STATE Initialize $\hat{x}_k(s,c)\leftarrow 0$ and $w(s,c)\leftarrow 0$ for all $s,c$ 
\FOR {$n\in\{1,2,\dotsc,n_\text{iter}\}$}
\STATE Find $\hat{\mathbf{X}}_k^*$ by solving ~\eqref{eq:problem_cvx}
\STATE Update $\displaystyle w(s,c)\leftarrow\beta/\left(\tau+\hat{x}(s,c)\right)$
\ENDFOR
\STATE $\mathbf{X}_k^*\leftarrow \round\left(\hat{\mathbf{X}}_k^*\right)$ 
\STATE Run Algorithm~\ref{alg:valid_allocation} on $\mathbf{X}_k^*$
\RETURN $\mathbf{X}_k^*$
\end{algorithmic}
\end{algorithm}
}

\subsection{Distributed Optimization}
Besides the global optimization presented in the previous section, we formulate a distributed optimization algorithm where the satellite-to-cell matching and resource allocation are solved separately. This algorithm, listed as pseudocode in Algorithm~\ref{alg:dist}, is used as a benchmark for the global optimization presented above.

As a first step, a maximum weighted cell-to-satellite matching problem is solved. That is, the cells are matched to the satellite that maximizes a given weight $w_k(s,c)$. As a result, a given satellite $s$ is matched to a set of cells 
\begin{equation}
\mathcal{C}_s=\left\{c\in\mathcal{C}:\argmax_{s'}w_k(s',c)\right\}, \quad \text{s.t. } \mathcal{C}_s\bigcap_{s'\in\mathcal{S}\setminus s}\mathcal{C}_{s'}=\emptyset.
\label{eq:cs}
\end{equation}
Afterwards, the resource allocation problem at satellite $s$ is formulated as the local optimization problem
\begin{IEEEeqnarray*}{rll}\label{eq:problem_local}
\hat{\mathbf{x}}_k^{(s)*}=\,& \argmax_{\hat{\mathbf{x}}_k^{(s)}}& \hspace{-0.8em}\sum_{c\in\mathcal{C}_s}\!U_c\log \Big(\hat{R}_k(s,c)\big(1-h_k(s,c)\big)\Big)\IEEEyesnumber\\[0.2em]
&\text{subject to}~&
\displaystyle0\leq \hat{x}_k(s,c)\leq N_T, \quad\forall s \in \mathcal{S}, c\in\mathcal{C}_s,\\[0.2em]
&&\displaystyle \sum_{c \in \mathcal{C}_s}\hat{x}_k(s,c)\leq N_TN_B, \quad\forall s \in \mathcal{S}, 
\end{IEEEeqnarray*}

Note that the objective function in~\eqref{eq:problem_local} is concave and its constraints are linear. Hence, it can be solved using standard convex programming methods and solvers.

\begin{algorithm}[t]
\caption{Algorithm for distributed optimization.}\label{alg:dist}
\begin{algorithmic}[1]
\footnotesize
\REQUIRE $\mathbf{X}_{k-1}$ and $\rho_k^\text{min}(s,c)$ for all $s,c$
\STATE Initialize  $\hat{x}_k(s,c)\leftarrow 0$ for all $s,c$

\FOR {$s\in\mathcal{S}$}
\STATE Find $\mathcal{C}_s$ as defined in~\eqref{eq:cs}
\STATE Find $\hat{\mathbf{x}}_k^{(s)}$ by solving~\eqref{eq:problem_local}
\ENDFOR
\STATE $\mathbf{X}_k^*\leftarrow \round\left(\left[\hat{\mathbf{x}}_k^{(1)},\hat{\mathbf{x}}_k^{(2)},\dotsc,\hat{\mathbf{x}}_k^{(S)}\right]\right)$ 
\STATE Run Algorithm~\ref{alg:valid_allocation} on $\mathbf{X}_k^*$
\RETURN $\mathbf{X}_k^*$
\end{algorithmic}
\end{algorithm}

\subsection{Performance Analysis}
We evaluate the performance of the network in terms of the throughput of the users and the amount of handovers.
Specifically, we consider both the actual throughput of the users and Jain's fairness index, given as follows.\vspace{-0.2cm}
\begin{equation}
    \mathcal{J}\left(\mathbf{X}_k\right)=\frac{\displaystyle\left(\sum_{c\in\mathcal{C}}U_cR_k\left(\mathbf{X}_k,c\right)\right)^2}{\displaystyle\left(\sum_{c\in\mathcal{C}}U_c\right)\sum_{c\in\mathcal{C}}U_c\left(R_k\left(\mathbf{X}_k,c\right)\right)^2}.\vspace{-0.2cm}
\end{equation}
Furthermore, the global optimization requires a number of iterations to reach a feasible solution, as it solves a dual problem in which cells can be connected to multiple satellites. The Lagrange multipliers $w(s,c)$ need to grow significantly before the problem is solved, and iterations can be computationally expensive, so we considered an \emph{early stopping} solution: Algorithm~\ref{alg:valid_allocation} can be used to find the point in the feasible region closest to the output of each iteration of the dual optimization, so that an approximate solution to the problem is reached. Naturally, the more iterations of the dual problem we can run before early stopping, the closer the approximation will be to the actual optimum, at the cost of a longer computation time. On the other hand, the distributed optimization does not need multiple iterations, and consequently has a much more predictable, and much lower, computational cost.

\section{Simulation Settings and Results}
\label{sec:results}

We consider a rectangular area covering central Europe (the same shown in Fig.~\ref{fig:map}), between latitudes $40^\circ$ and $55^\circ$ North  and longitudes $5^\circ$ and $30^\circ$ East. The cells are evenly spaced, each covering $0^\circ 15^\prime$ in both latitude and longitude and the population of each cell is obtained from ~\cite{CIESIN}. Therefore, there are a total of $6161$ cells in the area, out of which $766$ have zero users because they are over the sea or entirely unpopulated areas. Furthermore, we consider the $S=1584$ Starlink satellites deployed in the orbital shell with altitude $550$\,km. In our experiments, the number of satellites within range of the cells is $|S_k|\in\{19,20, \dotsc,25\}$ for all $k$. The communication parameters were taken from the specifications in the recent \gls{3gpp} technical reports on \glspl{ntn}~\cite{3GPPTR38.821}, and the full set of simulation parameters is given in Table~\ref{tab:params}.

\begin{table}[t]
    \centering
    \caption{Simulation parameters}\vspace{-0.2cm}
    {\renewcommand{\arraystretch}{1}
    \begin{tabular}{@{}lcc@{}}
    \toprule
    Parameter & Symbol & Value \\ \midrule
    \multicolumn{3}{c}{\textbf{Communication system}} \\ \midrule
    Carrier frequency (GHz) & $f$ & $2$\\ 
    Transmission power (W) &$P_\text{tx}$ & $75.35$\\
    Satellite antenna gain (dBi) & $G^{(s,c)}$ & $30$\\
    User antenna gain (dBi) & $G^{(c,s)}$& $0$\\
     Atmospheric loss (dB)  & $\theta_\text{dB}$ & $0.5$\\ 
    Pointing loss (dB)  & $\ell_\text{dB}$ & $3$\\ 
     System bandwidth (MHz)& $B$ & $30$\\
     Noise power (dBW) & $P_N$ & $-122.20$\\
     Duration of time slot (s) & $T$ &$10$\\
     Duration of \gls{ofdma} frame (ms) & $T_F$ & $10$\\
     Max. number of beams per satellite & $N_B$ & $10$\\
     Ratio of active users to total population& $\alpha$ & $1\times10^{-3}$\\ \midrule
     \multicolumn{3}{c}{\textbf{Satellite constellation}}\\ \midrule
     Total number of satellites & $S$ & $1584$\\
     Number of orbital planes & $P$ & $72$\\
     Altitude of deployment (km) & $h$ & $550$\\
     Inclination (deg) & $\delta$ & $53$\\
     Change in true anomaly between & $\Delta_\theta$ & $0$\\  satellites in adjacent orbital planes (deg)\\ 
     \bottomrule
    \end{tabular}\vspace{-0.4cm}}
    \label{tab:params}
\end{table}

The nominal data rates at each slot $\rho_k^\text{min}(s,c)$ were obtained through simulator coded in Python, which replicates the orbital dynamics of the constellation. A total of $100$ consecutive time slots were simulated. The optimization problems were solved using the CVXPY package~\cite{diamond2016cvxpy} using MOSEK ApS as solver. These were run on a PC with Windows 11 with an AMD Ryzen 7 5700U CPU at 1.8 GHz and with 64 GB of RAM.

\begin{figure}[t]
\includegraphics{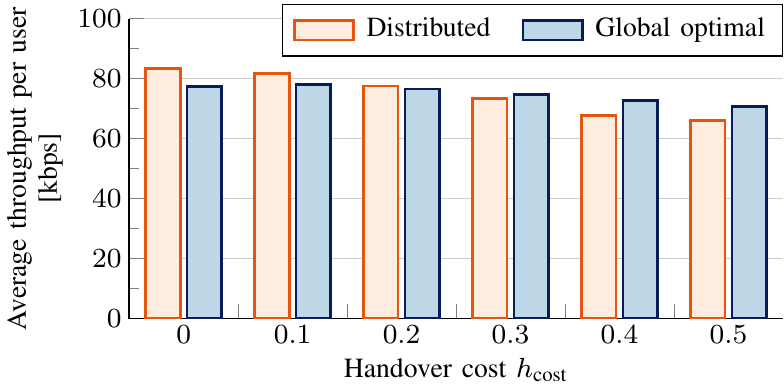}\vspace{-0.2cm}
\caption{Throughput per user with distributed allocation and global optimization for different values of the handover cost $h_{\text{cost}}$. }\vspace{-0.4cm}
\label{fig:avg_rates}
\end{figure}

We can analyze the average throughput per user within the area for different values of the handover cost $h_{\text{cost}}$. Intuitively, setting a higher value of the handover cost $h_{\text{cost}}$ aims to reduce the number of handovers and, thus, to achieve a more stable mapping between cells and satellites. However, a higher value of $h_{\text{cost}}$ may affect the efficiency of the system: if a cell remains matched to a satellite moving further away from it to avoid an expensive handover, it will experience a higher path loss due to the longer distance, and thus a lower throughput. This is confirmed by Fig.~\ref{fig:avg_rates}, which shows the average throughput per user for different values of $h_{\text{cost}}$, where the throughput for both the distributed and centralized algorithms decreases as $h_{\text{cost}}$ grows. This is because, with a higher value of $h_{\text{cost}}$, the algorithms focus more on stability and less on capacity maximization. 

Interestingly, the average throughput of the distributed allocation often matches that of the global optimization, and even outperforms it if the handover cost is low. the reason for this is that the global optimization aims for a load-balanced and fair solution, and can have a slightly smaller total throughput in order to achieve a fairer allocation of resources. Nevertheless, the average throughput decreases more rapidly with the distributed allocation than with the global optimization and, hence, the latter achieves a better balance between stability and performance than the distributed allocation.

\begin{figure}[t]
\includegraphics{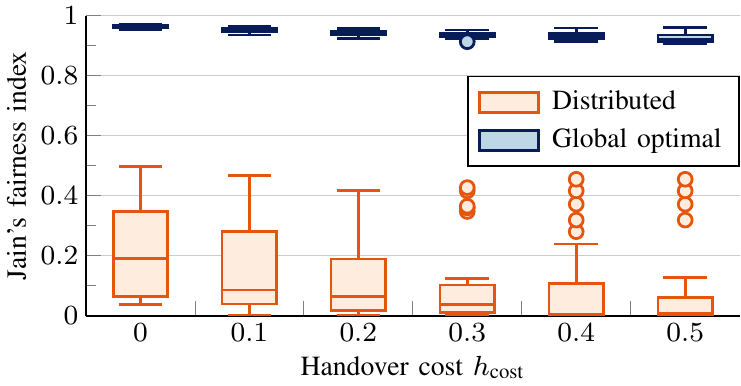}\vspace{-0.3cm}
\caption{Jain's fairness index among users at each realization of the allocation with distributed allocation and the global optimal for different values of the handover cost $h_{\text{cost}}$.}\vspace{-0.3cm}
\label{fig:fairness}
\end{figure}

Next, Fig.~\ref{fig:fairness} shows the distribution of Jain's fairness index for the two algorithms. Clearly, the global optimization achieves an exceedingly high fairness, maintaining the index above $0.9$ for all the considered values of the handover cost. In contrast, distributed allocation performs much worse, achieving a fairness index that never exceeds $0.5$. That is, the fairness index obtained with global optimization is, in most cases, more than two times higher than that of the distributed allocation. Furthermore, the fairness of the distributed algorithm decreases as the handover cost increases, as did its average throughput: for values of $h_{\text{cost}}>0.3$, the distributed strategy performs worse in terms of both average throughput and fairness than the global optimization. 

\begin{figure}[t]
\includegraphics{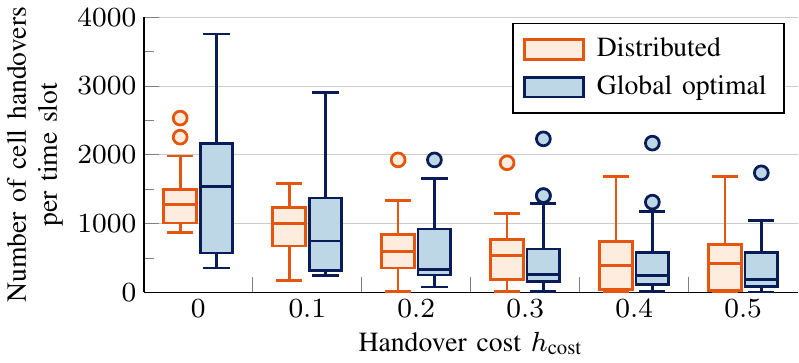}\vspace{-0.3cm}
\caption{Box plot of the number of handovers per time slot with distributed allocation and the global optimal for different handover costs $h_{\text{cost}}$.}\vspace{-0.4cm}
\label{fig:handovers}
\end{figure}

Fig.~\ref{fig:handovers} shows the distribution of the number of handovers per time slot as a function of the handover cost $h_\text{cost}$. As expected, the number of handovers decreases as $h_\text{cost}$ increases. Furthermore, the number of handovers with the global optimal algorithm is greater than that with the distributed allocation for $h_\text{cost}< 0.4$, but the opposite is true for $h_\text{cost}\geq 0.4$, where the number of handovers is reduced by more than $70\%$ when compared to $h_\text{cost}=0$. Thus, the global optimal algorithm is more efficient at reducing the number of handovers than the distributed allocation.

\begin{figure}[t]
    \centering
    \subfloat[]{\includegraphics{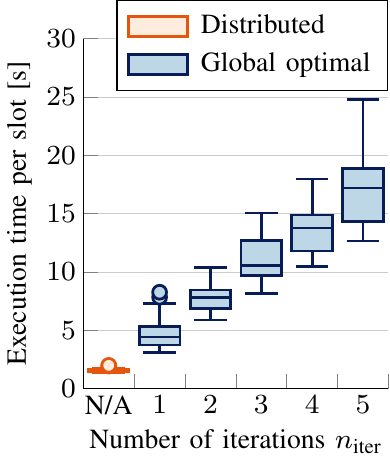}}\hfil
    \subfloat[]{\includegraphics{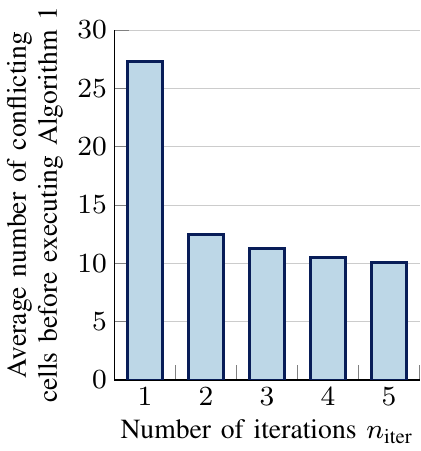}}
    \caption{(a) Execution time and (b) number of cells that violate the last constraint in~\eqref{eq:problem_x} as a function of the number of iterations $n_\text{iter}$ of Algorithm~\ref{alg:el1}.}\vspace{-0.4cm}
    \label{fig:exec_time}
\end{figure}

Finally, Fig.~\ref{fig:exec_time} shows the empirical results on execution time for the two algorithms, considering the number of iterations for the global optimization $n_\text{iter}$. Furthermore, Fig.~\ref{fig:exec_time} shows the average number of \emph{conflicting cells}, that is, the cells that are matched to more than one satellite and, hence, violate the last constraint in~\eqref{eq:problem_x} before the execution of Algorithm~\ref{alg:valid_allocation} as a function of $n_\text{iter}$. This latter metric is relevant because the cells that violate this constraint after the $n_\text{iter}$ iterations only receive a sub-optimal allocation of resources.

Clearly, the distributed strategy has a significant advantage over the centralized optimization in terms execution time and, besides, it guarantees that the cells are matched to exactly one satellite. Nevertheless, the execution time of the global optimization with $n_\text{iter}=\{1,2\}$ is below the selected time slot duration of $10$, which ensures that the global optimization can be executed in real-time in the selected computer architecture. Furthermore, the average number of conflicting cells is below $0.5\%$ of the total cells with users for $n_\text{iter}=1$ and it drops to just $0.2\%$ with $n_\text{iter}=2$. Even though the resource allocation for the conflicting cells is sub-optimal, the average number of conflicting cells is sufficiently low so that the differences in both the achieved average throughput and fairness index between $n_\text{iter}=1$ and $n_\text{iter}=5$ are negligible: less than $0.25\%$. Therefore, selecting a low $n_\text{iter}$ has a minor impact in the overall performance and its value can be selected to fulfil a specific real-time requirements for the given platform.

\section{Conclusions and Future Work}\label{sec:conc}

In this work, we designed two algorithms to optimize the allocation of downlink resources from \gls{leo} satellite networks to fixed cells on the ground. The algorithms were designed to optimize for fairness and efficiency in terms of per-user throughput, while minimizing the number of handovers. We evaluated the efficiency of dual-based optimization methods with respect to these three performance measures, showing that our approach can be tuned to achieve different balances in the trade-off between handover frequency and throughput while maintaining a high fairness. We also analyzed the computational complexity of the algorithms, proposing early stopping, which enables efficient real-time execution with a negligible performance loss.

Future directions on the optimization of resource allocation in \gls{leo} networks include joint uplink and downlink allocation and the optimization of resource allocation considering practical routing and higher layer mechanisms.

\bibliographystyle{IEEEtran}
\bibliography{IEEEabrv,bib}
\end{document}